\documentclass[12pt]{article}
\usepackage{amsmath}
\usepackage{amssymb}
\usepackage{array}
\usepackage{graphicx}
\newcommand{\pa}{\partial}
\newcommand{\comment}[1]{}
\begin{document}
\sloppy

\title{Spatially inhomogeneous states of charge carriers in graphene}
\author{A.\,V. Chaplik\thanks{e-mail: {\it chaplik@isp.nsc.ru}}, T.\,Ya. Tudorovskiy}
\date{}
\maketitle

\begin{center}
Institute of Semiconductor Physics, Siberian Branch of The Russian Academy of Sciences, \\
630090, Novosibirsk, Russia
\end{center}

\abstract{We study an interaction of 2D quasiparticles with linear
dispersion $E=\pm u|p|$ (graphene) with impurity potentials. It is
shown that in 1D potential well (quantum wire) there are discrete
levels, corresponding to localized states, whereas in 2D well
(quantum dot) there are no such states.  Scattering cross-section
of electrons (holes) of graphene by an axially symmetric potential
well is found and it is shown that for infinetily large energy of
 incoming particles  the cross-section tends to a constant.
 The effective Hamiltonian for a curved quantum wire of
graphene is derived and it is shown that the corresponding
geometric potential cannot form 1D bound states.}

\section{Introduction}

Monatomic layer of carbon atoms, forming hexagonal lattice
(graphene), is studied very intensively at present
\cite{NovScience}-\cite{NovNature}. ``Conical'' dispersion law for
quasiparticles (the name is borrowed from the similar 3D model of
a gapless semiconductor) results in crucial distinctions of their
dynamical characteristics from the corresponding characteristics
of massive particles. The density of electron states tends
linearly to zero as a function of energy $E$, counted from the
conical point, that is more rapidly  than for usual particles in
3D case ($\sqrt{E}$). This is a reason to expect that  formation
of bound states in potential wells will be hindered.

In the present paper we consider some simple exactly solvable
models of 1D and 2D potential wells from the viewpoint of
possibility to form bound states for  quasiparticles , described
by the Hamiltonian
\begin{equation}
\widehat{\mathcal{H}}=u\boldsymbol{\sigma \hat p},
\label{InHam}
\end{equation}
where $\boldsymbol{\sigma}=(\sigma_1,\sigma_2)$ are Pauli
matrices, $\boldsymbol{\hat p}=-i\hbar\nabla$ is momentum
operator, $u$ is characteristic velocity (for graphene $u\sim
10^6$ m/sec). It turned out, that quantum wire (1D localization)
is possible, whereas quantum dot as well as a hydrogen-like donor
(acceptor) are not possible. 1D effective Hamiltonian for a curved
wire will be derived and it will be shown that geometric
potential differs significantly from the case of parabolic
dispersion law.

Some problems of interaction of quasiparticles described by the Hamiltonian \eqref{InHam} with electrostatic potentials have recently been considered. In Ref.\,\cite{n-p-Falko} the transmission coefficient of carriers through 1D barrier (p-n junction) is found; same authors have shown in Ref.\,\cite{Friedel-Falko} that Friedel oscillations of the charge density around an impurity atom in graphene differs essentially from the ones in 2D electron systems with parabolic dispersion law. Bound states in a symmetric 1D potential well have been investigated in Ref.\,\cite{Pereira}. We  expound here solution of this problem (together with a more general one for an asymmetric well) as a starting point for our consideration of a curved  quantum wire.

\section{1D potential well}

The motion of electrons in a graphene waveguide, representing 2D stripe with straight axis, is described by the equation:
\begin{equation}
u(\boldsymbol{\sigma \hat p})\Psi+v(y)\Psi=E\Psi,
\label{massless2}
\end{equation}
where $v(y)$ is the potential confining the particle in the waveguide (below we suppose $u=\hbar=1$). Let us look for the solution in the form $\Psi(x,y)=\chi(y)\exp(ip x)$, where $\chi(y)=(\chi_1,\chi_2)$ is two-component spinor, whose components satisfy the equations:
\begin{eqnarray}
\left(-\frac{\pa}{\pa y}+p\right)\chi_2=(E-v(y))\chi_1 \quad
\left(\frac{\pa}{\pa y}+p\right)\chi_1=(E-v(y))\chi_2.
\label{eqn10}
\end{eqnarray}
After excluding $\chi_2$, we find:
\begin{equation}
\left[-\frac{\pa^2}{\pa y^2}+p^2+
\left(\frac{\pa}{\pa y}\ln(E-v(y))\right)\left(\frac{\pa}{\pa y}+p\right)\right]\chi_1=\left(E-v(y)\right)^2\chi_1
\end{equation}
Substituting $\chi_1$ in the form $\sqrt{E-v(y)}\widetilde\chi_1$, we obtain
\begin{equation}
\left[-\frac{\pa^2}{\pa y^2}+p^2-(E-v(y))^2+\frac{v''(y)/2-v'(y)p}{E-v(y)} +\frac{3(v'(y))^2}{4(E-v(y))^2}\right]\widetilde\chi_1=
0
\label{SchZero1}
\end{equation}
The equation for function $\chi_2=\sqrt{E-v(y)}\widetilde\chi_2$ is obtained by means of replacement $p\rightarrow -p$.

Let us consider the potential $v(y)$ in the form of step function:
$v(y)=0$ if $y<-a$, $v(y)=-v_0$ if $|y|<a$, $v(y)=-v_1$ if $y>a$.
In each region the equation \eqref{SchZero1} is reduced to
\begin{equation*}
-\pa^2\chi_1/\pa y^2=[(E-v_i)^2-p^2]\chi_1,
\end{equation*}
where $v_i=0$ if $y<-a$, $v_i=-v_{0,1}$ if $|y|<a$ and $y>a$ correspondingly. Solutions of the last equation decreasing when $y\to\pm\infty$, have the form $\chi_1=A_1 e^{\varkappa y}$ if $y<-a$, $\chi_1=B_1\sin qy + B_2\cos qy$ if $|y|<a$, $\chi_1=A_3 e^{-\varkappa_1 y}$ if $y>a$. Here $\varkappa=\sqrt{p^2-E^2}$, $q=\sqrt{(E+v_0)^2-p^2}$,
$\varkappa_1=\sqrt{p^2-(E+v_1)^2}$.

Matching conditions  $\chi_i|_{y=\pm a-0}=\chi_i|_{y=\pm a+0}$, $i=1,2$ lead to the following equation, determining the spectrum $E=E^\nu(p)$, where $\nu$ is the number of subband of transversal quantization:
\begin{eqnarray}
\left[1-\frac{\varkappa+p}{E}\ \frac{\varkappa_1-p}{E+v_1}+
\frac{p}{E+v_0}\ \frac{\varkappa_1-p}{E+v_1}-
\frac{\varkappa+p}{E}\frac{p}{E+v_0}\right]\sin(2qa)-\nonumber\\
-\frac{q}{E+v_0}\left[\frac{\varkappa_1-p}{E+v_1}+\frac{\varkappa+p}{E}\right]\cos(2qa)
=0.
\label{dtrm1}
\end{eqnarray}

In the case of symmetric well $v_1=0,\,\varkappa_1=\varkappa$;
equation \eqref{dtrm1} is simplified:
\begin{equation}
[E(E+v_0)-p^2]\sin(2qa)-\varkappa q\cos(2qa)=0.
\label{eqn29}
\end{equation}

Branches $E^\nu(p)$ for symmetric well are shown on Fig.1. Let us
note that equation \eqref{eqn29} always has the solution
$E(p)=|p|-v_0$, i.e. $q=0$. However this branch of the spectrum is
not physical since it corresponds to the wavefunction identically
equaled to zero.

\begin{figure}
\begin{center}
\scalebox{0.4}{\includegraphics[0,0][735,470]{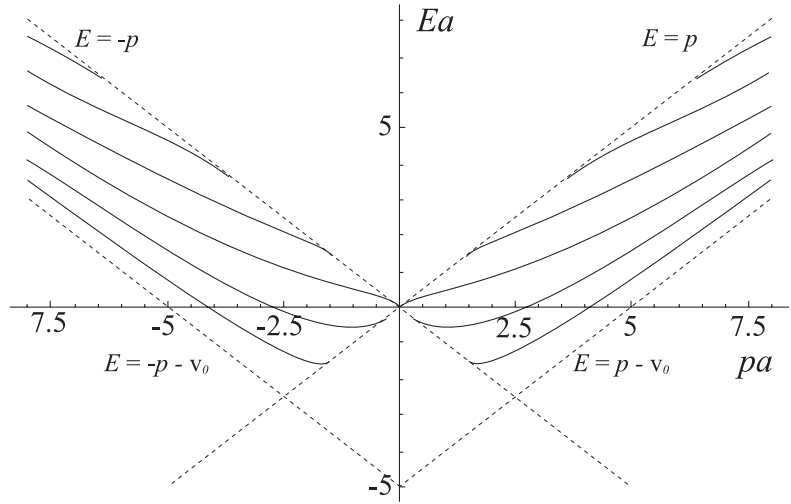}}
\end{center}
\caption{Bound states $E(p)$ in the 1D symmetric rectangular well; $v_0 a=5$. Dotted line shows the bounds of the area, occupied by discrete spectrum.}
\end{figure}

 Investigation of the equation \eqref{eqn29} leads to the condition
of reality of $q$, which, together with obvious condition of
reality of $\varkappa$, determines the region, occupied by
discrete spectrum $\max\{-|p|,|p|-v_0\}<E<|p|$. From this, in
particular, it follows that $E>-v_0/2$. It is easy to calculate
energy in the following limiting cases:
\begin{eqnarray}
  1) & |p|\ll v_0, \cos(2v_0 a)\neq 0: &
  E(p)=|p|{\rm sign}\{\sin(2v_0a)\}\cos(2v_0 a),
\label{Epsmall}\\
  2) & |p|\ll v_0, \cos(2v_0 a)=0: & E(p)=p^2/v_0, \\
  3) & v_0\to 0\ \ \textrm{èëè}\ \ a\to 0: & E=|p|(1-2v_0^2 a^2+\ldots).
\end{eqnarray}

In the shallow asymmetric well there are no bound states at any $p$. There are no bound states also in an asymmetric well of any depth at small enough $p$.

%The bound states in a symmetric 1D potential well have been recently found
%in the paper \cite{Pereira}. We expound here the solution of this
%problem as a starting point for consideration of a curved wire.

\section{Effective 1D equation in adiabatically curved stripe}

Let us consider a stripe with curved axis defined by the equation
$\boldsymbol{r}=\boldsymbol{R}(x);$ we suppose the stripe to be
curved in its plane. Here $\boldsymbol{R}(x)=(R_1(x),R_2(x))$ is a
smooth vector-function, $x$ is the natural parameter on the axis
(length counted out from some fixed point),
$|d\boldsymbol{R}/dx|=1$. Let
$\boldsymbol{n}(x)=\bigl(n_1(x),n_2(x)\bigr)$  be the unitary
vector normal to $d\boldsymbol{R}/dx$. In the neighborhood of axis
the curvilinear coordinates $x,\,y$, defined by the equality
$\boldsymbol{r}=\boldsymbol{R}(x)+y\boldsymbol{n}(x)$ can be
introduced. The equation \eqref{massless2} in curvilinear
coordinates has the form:
\begin{equation}
\left[\frac{(\boldsymbol{\sigma R}')}{\sqrt{1-ky}}(-i\frac{\pa}{\pa x})
\frac{1}{\sqrt{1-ky}}+
(\boldsymbol{\sigma n})(-i\frac{\pa}{\pa y})
-\frac{ik(\boldsymbol{\sigma n})}{2(1-ky)}
+v(y)\right]\widetilde\Psi=E\widetilde\Psi,
\label{curvlstripe2}
\end{equation}
where $k$ is the curvature of the axis of waveguide at the point $x$, $\widetilde\Psi=(1-ky)^{1/2}\Psi$. Let us take into account now that the stripe is curved adiabatically, i.e. formally $k\to 0$. Following the idea \cite{BerlyandDobr}, we look for the solution of the last equation in the form $\Psi^\nu(x,y)=\hat\chi^\nu\psi^\nu(x)$, $\hat\chi^\nu=(\hat\chi^\nu_1,\hat\chi^\nu_2)$ is two-component vector-operator and scalar wavefunction $\psi^\nu(x)$ satisfies effective Schr\"odinger equation
\begin{eqnarray}
\label{effSch}
%L^\nu(-i\frac{\pa}{\pa x},x)\psi^\nu(x)=E\psi^\nu(x),\quad
\hat L^\nu\psi^\nu(x)=E\psi^\nu(x).
\end{eqnarray}
Now we expand the  operator $\hat L$ in  powers of curvature $\hat
L^\nu=\hat H_{\rm eff}^\nu+\hat L_1^\nu+\ldots$. It will be shown
that operator $\hat H_{\rm eff}^\nu=H_{\rm eff}^\nu(-i\pa/\pa x)$
does not depend on $x$. It follows from \cite{BDT} that $H_{\rm
eff}^\nu(p)$ is an eigenvalue of the following problem:
\begin{equation}
\left[(\boldsymbol{\sigma R}')p
+(\boldsymbol{\sigma n})(-i\frac{\pa}{\pa y})+v(y)\right]\chi^\nu(p,x,y)=
H_{\rm eff}^\nu(p)\chi^\nu(p,x,y),
\label{terms}
\end{equation}
where $p$ is a parameter ($c$-number), $\chi^\nu=(\chi_1^\nu,\chi_2^\nu)$. Using the replacement $\chi_{2}^\nu=(n_2-in_1)\widetilde \chi_{2}^\nu$ we reduce \eqref{terms} to \eqref{eqn10}:
\begin{eqnarray}
\left[p-\frac{\pa}{\pa y}\right]\widetilde\chi_{2}^\nu=
\bigl(H_{\rm eff}^\nu(p)-v(y)\bigr)\chi_{1}^\nu,\quad
\left[p+\frac{\pa}{\pa y}\right]\chi_{1}^\nu=
\bigl(H_{\rm eff}^\nu(p)-v(y)\bigr)\widetilde\chi_{2}^\nu.
\label{chi2tilde}
\end{eqnarray}
It follows from the last equation that $H_{\rm
eff}^\nu(p)=E^\nu(p)$, i.e. $H_{\rm eff}^\nu$ does not depend on
$x$. Let us choose functions
$\chi_{1}^\nu,\,\widetilde\chi_{2}^\nu$ to be real. Then from
general formulae \cite{BDT} and the relation
$\left\langle(\chi^\nu)^+(\boldsymbol{\sigma
n})\chi^\nu\right>_y=0$ it follows that
\begin{equation}
\hat L_1^\nu=\left[k(x),\bigl(2K(\hat p)\hat p+Q(\hat p)\bigr)\right]_+, \quad
K(p)=\left\langle\chi_1^\nu y\widetilde\chi_2^\nu\right>_y, \quad Q(p)=\left\langle\chi_1^\nu
\widetilde\chi_2^\nu\right>_y.
\label{L1graphene2}
\end{equation}
Here $\hat p=-i\pa/\pa x$, $[\cdot,\cdot]_+$ means anticommutator,
and $\left<\cdot\right>_y$ means integration over $y$. For even
potential $K(p)=0$.

{\bf Effective equation for rectangular well at small $|p|$.}
For symmetric rectangular well at small $|p|$ one can use dispersion relation \eqref{Epsmall}. Then we find $Q(0)={\rm sign}\{\sin(2v_0a)\}{\rm sign}\{\hat p\}\cos(2v_0a)/2$. Effective 1D equation for small momentums takes the form:
\begin{equation}
{\rm sign}\{\sin(2v_0a)\}{\rm sign}\{\hat p\}
\left(-i\frac{\pa}{\pa x}+\frac{k(x)}{2}\right)\psi=E\psi
\label{effSch-bottom}
\end{equation}
Thus, the geometric potential for quasiparticles has the form
$k(x)/2$ and, how one can see from \eqref{effSch-bottom}, it
cannotform bound states.

\section{Axially symmetric potential}

Let us consider the equation \eqref{massless2} for the axially-symmetric potential. In the cylindric coordinates $x=r\cos\varphi,\,y=r\sin\varphi$ it has the form
\begin{eqnarray}
e^{-i\varphi}\left(-i\frac{\pa}{\pa r}-\frac{1}{r}\frac{\pa}{\pa\varphi}\right)\Psi_2=(E-v(r))\Psi_1,\quad
e^{i\varphi}\left(-i\frac{\pa}{\pa r}+\frac{1}{r}\frac{\pa}{\pa\varphi}\right)\Psi_1=(E-v(r))\Psi_2.
\end{eqnarray}
We find the solution in the form $\Psi_1=e^{in\varphi}\chi_1,\,\Psi_2=e^{i(n+1)\varphi}\chi_2$, where $\chi_1,\,\chi_2$ satisfy equations
\begin{eqnarray}
\label{chi1}
\left(-i\frac{\pa}{\pa r}-\frac{i(n+1)}{r}\right)\chi_2=(E-v(r))\chi_1,\quad
\left(-i\frac{\pa}{\pa r}+\frac{in}{r}\right)\chi_1=(E-v(r))\chi_2.
\label{chi2}
\end{eqnarray}

\subsection{Rectangular well}

Let us consider axially symmetric well with constant width: $v(r)=-v_0$ if $r<a$ and $v(r)=0$ if $r\geq a$. Excluding $\chi_2$ from \eqref{chi2}, we obtain:
\begin{eqnarray}
\left\{
-\frac{\pa^2}{\pa r^2}-\frac{1}{r}\frac{\pa}{\pa r}+\frac{n^2}{r^2}
-(E+v_0)^2\right\}\chi_1=0,\quad r<a,\\
\left\{
-\frac{\pa^2}{\pa r^2}-\frac{1}{r}\frac{\pa}{\pa r}+\frac{n^2}{r^2}
-E^2\right\}\chi_1=0, \quad r\geq a.
\label{26}
\end{eqnarray}
These equations have solutions $J_n(|E+v_0|r),\,N_n(|E+v_0|r)$ and
$J_n(|E|r),\,N_n(|E|r)$ correspondingly. The solution regular in
the point $r=0$ has the form $\chi_1(r<a)=C_n J_n((E+v_0)r)$.

The equation \eqref{26} contains only the square of the energy.
Thus its solutions don't depend on the sign of $E$. These
solutions are equivalent to scattering states of the usual radial
Schr\"odinger equation with $E>0$. From this circumstance one can
conclude that there are no bound states in such a potential well.
We would like to emphasize that this conclusion does not depend on
the depth and width of the well, i.e. 2D localization of
quasiparticals in graphene (quantum dot) is principally impossible
(naturally, this statement only relates to the region of momenta
where the Hamiltonian \eqref{massless2} and linear dispersion law
are valid). The same is true, obviously, for any potential
decreasing at the infinity, hence, hydrogen-like states of donors
or acceptors in graphene do not exist.

Let us consider now the scattering problem. Let the wave with
positive energy comes from the infinity along the $x$-axis. Then
the wavefunction $\Psi$ far from the origin has the form:
\begin{eqnarray}
\Psi\simeq\frac{ e^{i|E|r\cos\varphi}}{\sqrt{2}}
\begin{pmatrix}
1\\ 1
\end{pmatrix}
+ \frac{e^{i|E|r}}{\sqrt{r}}
\begin{pmatrix}
f_1(\varphi) \\ f_2(\varphi)
\end{pmatrix}.
\end{eqnarray}

Using the expansion of the plane wave
$\exp(i|E|r\cos\varphi)=({1}/{2})\sum_{n=-\infty}^\infty i^n
[H_n^{(1)}(|E|r)+H_n^{(2)}(|E|r)]\exp(in\varphi),$ we can
represent the solution in the form
\begin{equation}
\Psi=
\frac{1}{2}
\sum_{n=-\infty}^\infty
\begin{pmatrix}
i^n e^{in\varphi}(S_n^{(1)}H_n^{(1)}(|E|r)+ 2^{-1/2} H_n^{(2)}(|E|r)) \\
i^n e^{in\varphi}(S_n^{(2)}H_n^{(1)}(|E|r)+2^{-1/2} H_n^{(2)}(|E|r))
\end{pmatrix}.
\end{equation}
Hence
\begin{eqnarray}
f_1(\varphi)=%\frac{1}{\sqrt{2\pi k}}\sum_{n=-\infty}^\infty i^n e^{in\varphi}(S_n^{(1)}-\xi_1)e^{-i\pi n/2-i\pi/4}=
\frac{e^{-i\pi/4}}{\sqrt{2\pi |E|}}\sum_{n=-\infty}^\infty e^{in\varphi}(S_n^{(1)}- 2^{-1/2}),\nonumber\\
f_2(\varphi)=%\frac{1}{\sqrt{2\pi k}}\sum_{n=-\infty}^\infty i^n e^{in\varphi}(S_n^{(2)}-\xi_2)e^{-i\pi n/2-i\pi/4}=
\frac{e^{-i\pi/4}}{\sqrt{2\pi |E|}}\sum_{n=-\infty}^\infty e^{in\varphi}(S_n^{(2)}-2^{-1/2})
\end{eqnarray}

Withfunctional relations for Hankel functions, it is easy to show
that $S_{n+1}^{(2)}=S_n^{(1)}$. Hence
$f_2(\varphi)=e^{i\varphi}f_1(\varphi).$ Continuity conditions for
the wavefunction at $r=a$ lead to the equations:
\begin{eqnarray}
S_n^{(1)}H_n^{(1)}(|E|a)+2^{-1/2} H_n^{(2)}(|E|a)= C_n J_n(|E+v_0|a),\\
S_{n}^{(1)}H_{n+1}^{(1)}(|E|a)+2^{-1/2} H_{n+1}^{(2)}(|E|a)=
C_n J_{n+1}(|E+v_0|a)
\end{eqnarray}
From these equations one can find $S_{n}^{(1)}$, $S_{n}^{(2)}$.
Wavefunction of the scattered particles has the form
$\Psi_{out}\simeq{e^{i|E|r}}r^{-1/2} (f_1(\varphi),
f_2(\varphi)),$ and the current density is given by
$\boldsymbol{j}_{out}=\Psi_{out}^+\boldsymbol{\sigma}\Psi_{out}=
{2|f_1(\varphi)|^2}r^{-1} \{\cos\varphi,\sin\varphi\}$ (we write
Cartesian components of the current in braces). Thus, the
differential cross-section is ${d\sigma}/{d\varphi}=
2\,|f_1(\varphi)|^2.$ Total cross-section is
\begin{eqnarray}
\sigma=2\int_0^{2\pi}|f_1(\varphi)|^2d\varphi=
\frac{1}{|E|}\sum_{n=-\infty}^\infty |S_n+1|^2=\nonumber\\
=\frac{8}{|E|}\sum_{n=0 }^\infty \left|
\frac{J_n(|E+v_0|a)J_{n+1}(|E|a)-J_{n+1}(|E+v_0|a)J_n(|E|a)}
{J_n(|E+v_0|a)H_{n+1}^{(1)}(|E|a)-J_{n+1}(|E+v_0|a)H_n^{(1)}(|E|a)}
\right|^2.
\end{eqnarray}
Similar  calculations lead to the same formula for the total
cross-section for particles with negative energy. In the
low-energy limit $|E|a\ll 1$ the  asymptotic formulae
$J_0(z)=1+O(z^2),\,J_1(z)=z/2+O(z^2),\,N_0(z)={\rm
const}+2\ln(z)/\pi+O(z),\,N_1(z)=-2/(\pi z)+O(z^0)$ result in
\begin{equation} \sigma\simeq 2\left(\frac{\pi J_1(v_0a)}{J_0(v_0
a)}\right)^2 |E|a^2
\end{equation}

In Fig.2 we show the scattering cross-section as a function of
$|E|a$. One can see that cross-section has resonances. We will
study below the opposite limiting case $|E|a\gg 1$ for which a
different method is more convenient.

\begin{figure}
\begin{center}
\scalebox{1.2}{\includegraphics[0,0][255,189]{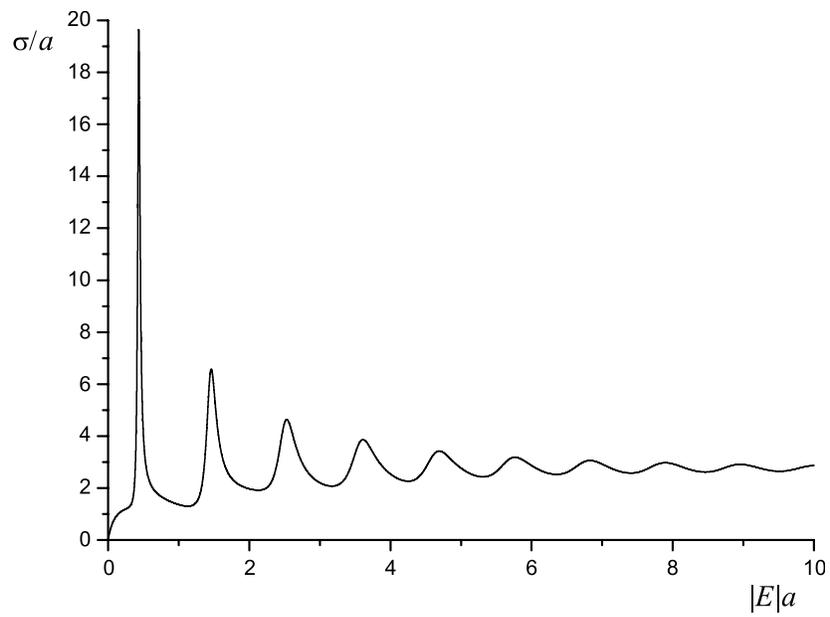}}
\end{center}
\caption{Scattering cross-section by rectangular axially symmetric well with radius $a$, $v_0 a=\pi$.}
\end{figure}

\subsection{Green function and integral  equation for the scattering problem}
Green function of the operator \eqref{InHam} is $2\times 2$-matrix $G(\boldsymbol{r})$. It satisfies the equation
\begin{equation}
[(\boldsymbol{\sigma \hat p})-E]G(\boldsymbol{r};E)=\delta(\boldsymbol{r})I,
\qquad G=\begin{pmatrix}
g_{11} && g_{12} \\
g_{21} && g_{22}
\end{pmatrix}, \qquad
I=\begin{pmatrix}
1 && 0\\
0 && 1
\end{pmatrix}.
\end{equation}

Solving this equation we find
\begin{equation}
G(\boldsymbol{r};E)=\frac{1}{4}\begin{pmatrix}
iEH_0^{(1)}(|E|r) && |E|e^{-i\varphi}\bigl(H_0^{(1)}\bigr)'(|E|r) \\
|E|e^{i\varphi}\bigl(H_0^{(1)}\bigr)'(|E|r) && iEH_0^{(1)}(|E|r)
\end{pmatrix}.
\end{equation}
For large $|E|r$ Green function has the asymptotic form
\begin{equation}
G(\boldsymbol{r};E)\simeq \sqrt{\frac{|E|}{8\pi r}}
e^{ i\left(|E|r+{\pi}/{4}\right)}
\begin{pmatrix}
{\rm sign}(E) && e^{-i\varphi} \\
e^{i\varphi} && {\rm sign}(E)
\end{pmatrix}.
\end{equation}

The scattering is described  by the integral  equation
\begin{equation}
\Psi(\boldsymbol{r};E)=\Psi_{in}(\boldsymbol{r};E)
-\int G(\boldsymbol{r}-\boldsymbol{r'};E)v(\boldsymbol{r'})\Psi(\boldsymbol{r'};E)
d^2\boldsymbol{r'},
\end{equation}
where $\Psi_{in}$ is the wavefunction of incoming particles. In
the first Born approximation we obtain:
\begin{eqnarray}
\Psi(\boldsymbol{r};E)=
\Psi_{in}(\boldsymbol{r};E)
-\frac{v_0a}{2q}\sqrt{\frac{\pi |E|}{ r}}e^{ i\left(|E|r+{\pi}/{4}\right)}
\begin{pmatrix}
1+e^{-i\varphi}\\
{\rm sign}(E)\,(1+e^{i\varphi})
\end{pmatrix}
J_1(qa),
\end{eqnarray}
where $q=2|E|\sin(\varphi/2)$.  Scattering cross-section is
\begin{eqnarray} \frac{d\sigma}{d\varphi}=\frac{2\pi |E|
v_0^2a^2}{q^2}\cos^2(\varphi/2) J_1^2(qa)= \frac{\pi
v_0^2a^2}{2|E|}\cot^2(\varphi/2)
J_1^2\bigl(2|E|a\sin(\varphi/2)\bigr),\nonumber\\
\sigma=\frac{\pi^2 v_0^2a^4 |E|}{2} \ {}_2 F_3[1/2, 3/2; 2, 2, 3;
-4|E|^2a^2],
\end{eqnarray}
where ${}_2 F_3$ is the generalized hypergeometric function.

When $|E|a\to \infty$ scattering cross-section tends to the
constant $\sigma\to {16 v_0^2a^3}/{3}$. Note that for usual
particles the Born scattering cross-section by a short-range
potential tends to zero when energy increases \cite{LL3}, however
if one formally supposes the mass to be proportional to momentum
(to obtain the linear dispersion law), then limiting value of the
Born cross-section at $|E|\to\infty$ also turns out to be
constant.

\vspace{1cm} \textit{Acknowledgements:-} The authors would like to
thank M.\,V. Entin and V.\,M. Kovalev for useful discussions. This
work has been  supported by THE RFBR (grant no. 05-02-16939), The
Council of The President of The RF for scientific schools (gr.
4500, 2006.2) and by the programs of The Russian Academy of
Sciences.

\end{document}